\documentclass[letter]{aa}

\usepackage{graphicx}
\usepackage{txfonts}
\usepackage{natbib}
\usepackage{multirow}
\usepackage{longtable}
\usepackage{subfig}

\begin{document}

\titlerunning{Discovery of magnetic fields in the very young, massive\\
stars W601 (NGC~6611) and OI~201 (NGC~2244)}

\title{Discovery of magnetic fields in the very young, massive\\
stars W601 (NGC~6611) and OI~201 (NGC~2244)
\thanks{Based on observations obtained at the Canada-France-Hawaii Telescope (CFHT) which is operated by the National Research Council of Canada, the Institut National des Sciences de l'Univers of the Centre National de la Recherche Scientifique of France, and the University of Hawaii}}

\author{
        E.\,Alecian \inst{1,} \inst{2}\thanks{visiting astronomer at the Pic du Midi Telescope Bernard Lyot (USR5026) of Universit\'e Paul Sabatier and CNRS Institut National des Sciences de l'Univers}
        \and
        G.A.\,Wade \inst{1}
        \and
        C.\,Catala \inst{2}
        \and
        S.\,Bagnulo \inst{3}
        \and
        T.\,Boehm \inst{4}
        \and
        D.\,Bohlender \inst{5}
        \and
        J.-C.\,Bouret \inst{6}
        \and
        J.-F.\,Donati \inst{4}
        \and
        C.P.\,Folsom \inst{3}
        \and
        J.\,Grunhut \inst{1}
        \and
        J.D.\,Landstreet \inst{7}
       }

\offprints{E. Alecian}

\institute{
           Dept. of Physics, Royal Military College of Canada, 
           PO Box 17000, Station Forces, Kingston, ON K7K 7B4, Canada\\
           \email{evelyne.alecian@rmc.ca}
           \and
           Observatoire de Paris, LESIA, 
           Place Jules Janssen, F-92195 Meudon Cedex, France
           \and
           Armagh Observatory, 
           College Hill, Armagh BT61 9DG, Northern Ireland, UK
           \and
           Laboratoire d'Astrophysique, Observatoire Midi-PyrŽnŽes, 
           14 avenue Edouard Belin, F-31400 Toulouse, France
           \and
           National Research Council of Canada, Herzberg Institute of Astrophysics, 
           5071 West Saanich Road, Victoria, BC V9E 2E7, Canada
           \and
           Laboratoire d'Astrophysique de Marseille, Traverse du Siphon, 
           BP8-13376 Marseille Cedex 12, France
           \and
           Dept. of Physics \& Astronomy, University of Western Ontario,
           ON N6A 3K7, London, Canada
          }

\date{Received November 27, 2007; accepted November 27, 2007}

%
\abstract
{Recent spectropolarimetric observations of Herbig Ae/Be stars have yielded new arguments in favour of a fossil origin for the magnetic fields of intermediate mass stars.}
{To study the evolution of these magnetic fields, and their impact on the evolution of the angular momentum of these stars during the pre-main sequence phase, we observed Herbig Ae/Be members of young open clusters of various ages.}
{We obtained high-resolution spectropolarimetric observations of Herbig Ae/Be stars belonging to the young open clusters NGC~6611 ($<6$~Myr), NGC~2244 ($\sim1.9$~ Myr), and NGC~2264 ($\sim8$~Myr), using ESPaDOnS at the Canada-France-Hawaii Telescope.}
{Here we report the discovery of strong magnetic fields in two massive pre-main sequence cluster stars. We detected, for the first time, a magnetic field in a pre-main sequence rapid rotator: the $10.2~M_{\odot}$ Herbig B1.5e star W601 (NGC~6611; $v\sin i\simeq 190$~km/s). Our spectropolarimetric observations yield a longitudinal magnetic field larger than 1~kG, and imply a rotational period shorter than 1.7 days. The spectrum of this very young object (age~$\sim 0.017$~Myr) shows strong and variable lines of He and  Si. We also detected a magnetic field in the $12.1~M_\odot$ B1 star OI~201 (NGC~2244; $v\sin i=23.5$~km/s). The Stokes V profile of this star does not vary over 5 days, suggesting a long rotational period, a pole-on orientation, or aligned magnetic and rotation axes. OI~201 is situtated near the Zero-Age Main Sequence on the HR diagram, and exhibits normal chemical abundances and no spectrum variability.}
{}

\keywords{}

\maketitle

%

\section{Introduction}

About 5\% of main sequence (MS) A and B stars host strong, organised magnetic fields. All these magnetic stars also show peculiar and variable chemical abundances,  and are called the magnetic Ap/Bp stars. The origin of these magnetic fields is still not well understood. The favoured hypothesis is the fossil field model, which proposes that the magnetic fields are relics of fields present in the molecular clouds from which these stars formed. This implies that some pre-main sequence (PMS) stars of intermediate mass should also be magnetic, with field properties similar to those of the Ap/Bp stars. Many such PMS stars are identifiable observationally as Herbig Ae/Be (HAeBe) stars.

Furthermore, most of the Ap/Bp stars rotate very slowly compared to "normal" (nonmagnetic and nonpeculiar) A/B stars. Their rotational periods 
are typically about 3~days, whereas normal A/B stars rotate with periods less that one day. Remarkably, some Ap/Bp stars have significantly longer rotational periods, which in extreme cases can approach one century.  Stepien (2000) considered different hypothesis to understand the origin of the slow rotation of Ap/Bp stars. He concluded that only magnetic braking, acting during the PMS phase of stellar evolution, could explain the slow rotation of the Ap/Bp stars. If Stepien's conclusion is correct then some HAeBe stars should host both magnetic fields and disks, and an evolution of the rotation of these stars should occur during their PMS phase.

Our recent studies of the magnetic fields of HAeBe stars have led to the discovery of 4 new magnetic HAeBe stars, bringing strong, new arguments in favour of the fossil-field hypothesis \citep[][Folsom et al. in prep.]{wade05,catala07,alecian08}.

To proceed further, we have begun to study the evolution of magnetism and rotation of intermediate-mass stars by observing HAeBe stars in young open clusters of different ages. We have obtained ESPaDOnS spectropolarimetric observations of a sample of 61 young intermediate-mass stars in the open clusters NGC~2244, NGC~2264, and NGC~6611. While a full description of the experiment and the results will be provided in a forthcoming paper, in this Letter we report the detection of magnetic fields in two young, massive stars: W601 (= NGC~6611~601) and OI~201 (=NGC~2244~201).

%

\section{Observations}

\begin{table}[t]
\caption{Log of observations of W601 and OI 201. Columns 1 and 2 give the UT date and the Heliocentric Julian Date of the observations. Column 3 gives the total exposure time. Column 4 gives the peak signal to noise ratio per 1.8~km/s CCD pixel (at $\sim708$~nm for W601, and at $\sim552$~nm for OI~201) in the spectra, and Column 5 gives the signal to noise ratio per 1.8~km/s in the LSD Stokes $V$ profiles. Column 6 gives the longitudinal magnetic field.}
\label{tab:log}
\centering
\begin{tabular}{@{\,}lcccc@{\,\,}r@{$\pm$}l@{\,}}
\hline\hline
Date (d/m/y)       & HJD               & $t_{\rm exp}$ & S/N & S/N     & \multicolumn{2}{c}{$B_{\ell}$} \\
UT Time & (2 450 000+) & (s)                  &        & (LSD)  & \multicolumn{2}{c}{(G)}           \\
\hline
\multicolumn{7}{c}{W601} \\
 10/08/06   6:09 & 3957.76004 & 3600 & 220 & 1910 &   -514 & 335 \\
 06/03/07 14:28 & 4166.10187 & 3600 & 180 & 1610 &   -313 & 427 \\
 06/03/07 15:35 & 4166.14879 & 3600 & 120 & 1010 & -1279 & 696 \\
 09/03/07 14:34 & 4169.10634 & 3600 & 190 & 1760 &  1391 & 394 \\
\\
\multicolumn{7}{c}{OI 201} \\
 04/03/07   9:49 & 4163.91136 &   960 & 150 & 1600 & -477 & 73 \\
 05/03/07   6:12 & 4164.76045 &   960 & 160 & 1820 & -543 & 65 \\
 06/03/07   6:47 & 4165.78497 &   960 & 160 & 1710 & -546 & 68 \\
 09/03/07   9:03 & 4168.87910 &   960 & 150 & 1490 & -471 & 76 \\
 \hline
\end{tabular}
\end{table}

We observed W601 and OI~201 in August 2006 and March 2007, using the ESPaDOnS spectropolarimeter installed at the Canada-France-Hawaii Telescope (CFHT, Hawaii). The observing and reduction procedures are fully described in many papers \citep[e.g.,][]{alecian08} and provide us with the intensity $I$ and circular polarisation $V$ spectra of a star, with a resolving power of about 65000 and S/N ratios ranging from 120 to 220 per CCD pixel for W601, and from 140 to 160 per CCD pixel for OI~201. The Libre-Esprit reduction package, developed especially for ESPaDOnS, calculates, in addition to the $I$ and $V$ spectra, a null polarisation spectrum $N$ \citep{donati97}, which allows us to check that the signature in the Stokes $V$ spectrum is not spurious. The log of the observations is reported in Table 1. 

To increase the effective signal to noise (S/N) ratio of our data, we applied the Least Squares Deconvolution \citep[LSD;][]{donati97} procedure using tailored line masks of 500 and 680 lines for W601 and OI~201, respectively. The S/N ratios of the LSD Stokes $V$ profiles are at least 10 times larger than the S/N ratio in the original spectra (Table \ref{tab:log}), and allow us to clearly see strong Zeeman signatures in each of our observations. 

\begin{figure}
\centering
\includegraphics[width=6cm,angle=90]{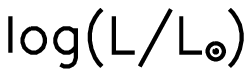}
\caption{A pre-main sequence Hertzsprung-Russell diagram showing the positions of W601 ({\it red square}) and OI~201 ({\it green diamond}). The CESAM PMS evolutionary tracks ({\it black full line}), the zero-age main sequence (ZAMS) ({\it dot-dashed line}), and the birthline ({\it blue dashed line}) are also shown.}
\label{fig:hr}
\end{figure}

%

\section{W601 (NGC~6611)}

\begin{figure}[t]
\centering
\includegraphics[width=6.5cm,angle=90]{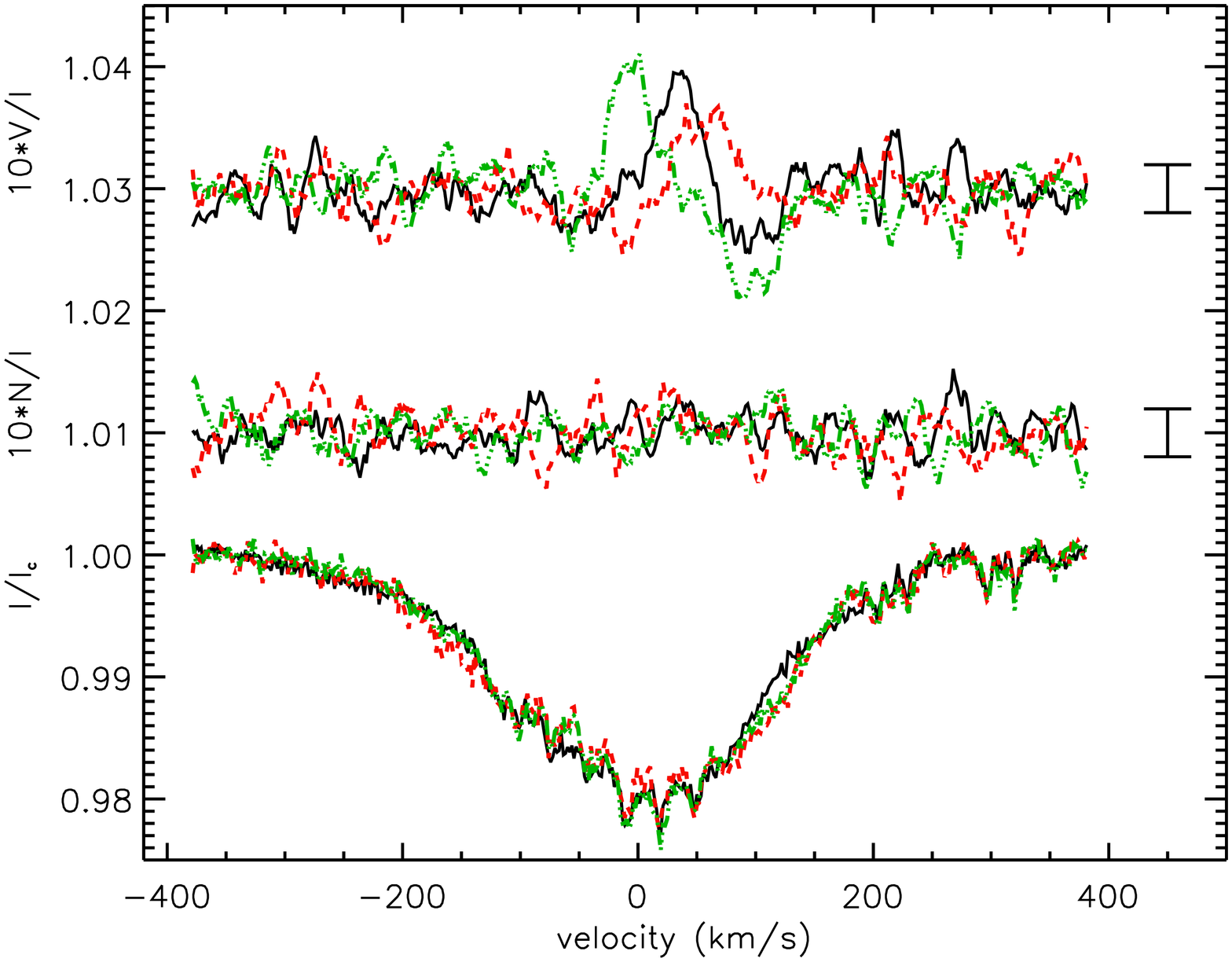}
\hfill
\includegraphics[width=5.6cm,angle=90]{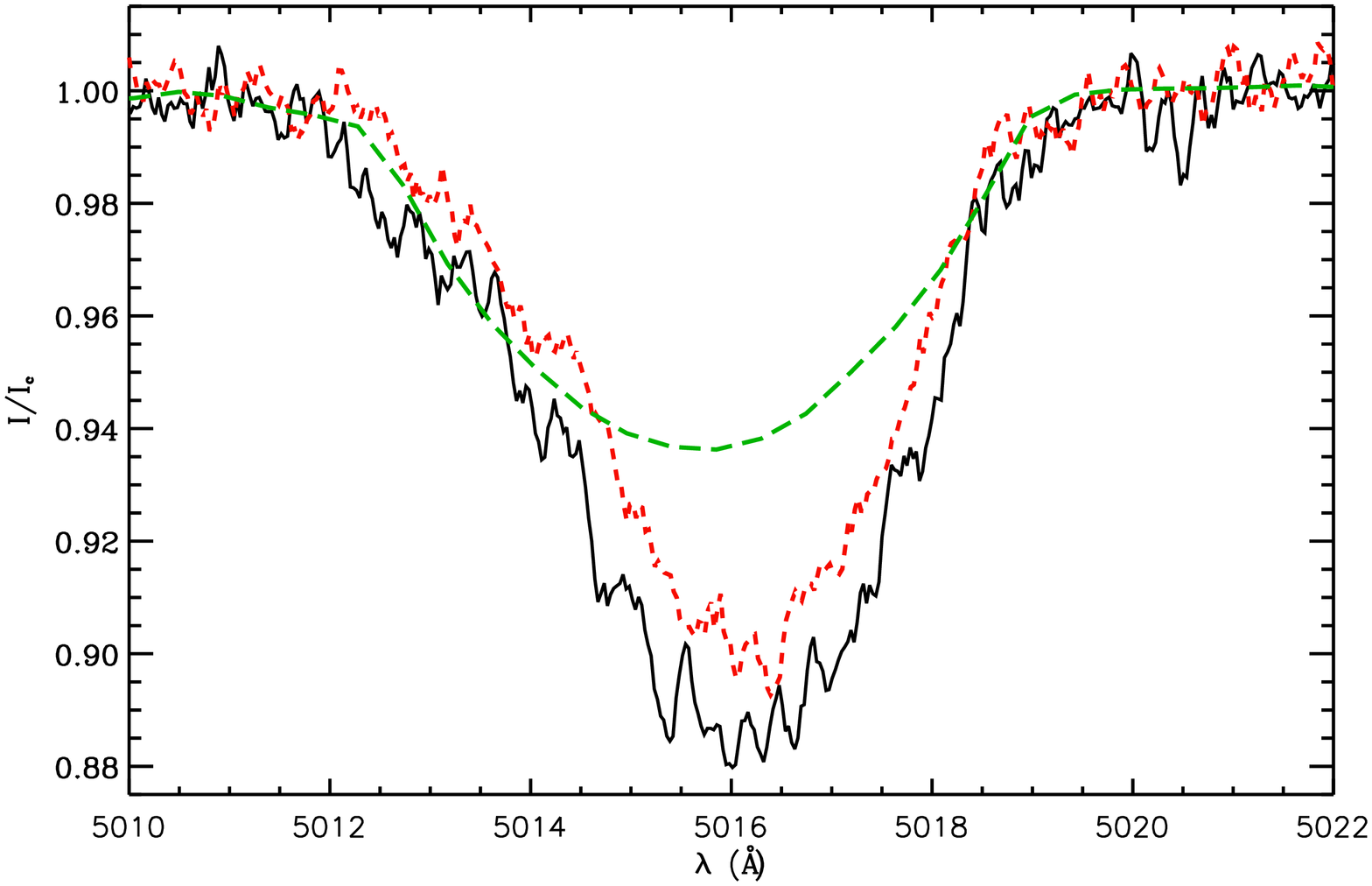}
\caption{{\em Upper frame:}\ LSD $I$ ({\it bottom}), $V$ ({\it upper}) and $N$ ({\it middle}) profiles of W601. {\it Full black line}: 10/08, {\it red-dashed line}: 6/03 14:28, {\it green dot-dot-dot-dashed line}: 9/03. The mean error bars are plotted next to each profile. {\em Lower frame:}\ The broad, variable He~{\sc i}\ $\lambda 5015$ of W601, observed on 6/03 ({\it full-black line}) and 9/03 ({\it red-dashed line}). The green long-dashed-line is a TLUSTY/SYNSPEC synthetic profile computed for a solar He abundance.}
\label{fig:w601}
\end{figure}

The B1.5 star W601 is member of the open cluster NGC~6611, situated at a distance of 1.8 kpc from the sun \citep{dufton06}, and has been classified as a PMS star by 
\citet{martayan07}. Due to scatter in the colour-magnitude diagram, the age of the cluster is not well constrained. \citet{dewinter97} conclude that the sample of PMS stars is not coeval, with ages spread from 0 to 6 Myr. According to the coordinates of the center of the cluster of \citet{belikov99} the position of W601 is offset, while still situated inside the limits of the cluster. Its proper motion from the Tycho-2 catalogue \citep{hog00}, its reddening law \citep{guarcello07}, and its position in the HR diagram \citep{dufton06} are fully consistent with NGC~6611 membership. 

According to \citet{kumar04},
this star exhibits mid-infrared (MIR) excess. In our data, we observe emission in  H$\alpha$ and H$\beta$. Given its membership in NGC~6611, its  HR diagram position, and its spectral properties, we concur with the classification of W601 as a PMS Herbig Be star.

Using the effective temperature ($\log T_{\rm eff}=4.35\pm0.05$) and luminosity ($\log L/L_{\odot}=3.96\pm0.11$) of \citet{dufton06}, we compared the HR diagram position of W601 with PMS evolutionary tracks of solar abundances, calculated using the CESAM code \citep{morel97}. We find a mass $M=10.2^{+1.2}_{-0.7}$~M$_{\odot}$ and a radius $R=6.4^{+2.7}_{-1.9}$~R$_{\odot}$, leading to $\log g=3.8^{+0.4}_{-0.3}$~(cgs). The age of the star has been estimated to be about $0.016^{+0.037}_{-0.013}$~Myr from the birthline calculated with a protostellar mass accretion rate of $10^{-4}$~M$_{\odot}$.yr$^{-1}$ \citep{palla92} (Fig. \ref{fig:hr}).

We obtained spectra of W601 sampling a variety of timescales, from one hour to several months (Table \ref{tab:log}). We observe emission in the cores of the H$\alpha$ and H$\beta$ Balmer lines, and in the wings of H$\alpha$, varying on a timescale of one hour. The absorption lines are strongly broadened, with ${\rm v}\sin i=190\pm10$~km.s$^{-1}$. According to our determination of the stellar radius, this implies that the rotation period of W601 is shorter than 1.7~d.

Most of the spectrum of W601 is consistent with a synthetic spectrum of $T_{\rm eff}=22500$~K, $\log g=4$~(cgs), broadened by ${\rm v}\sin i=190$~km.s$^{-1}$, calculated using TLUSTY non-LTE atmosphere models and the SYNSPEC code \citep{hubeny88,hubeny92}. However, we observe that all He~{\sc i} lines are substantially stronger than the synthetic ones calculated using the solar abundance (Fig. \ref{fig:w601}, lower frame). The He lines alos show variability in their depth and shape on a timescale of 1 hour. We also observe strong and variable Si~{\sc iii} lines, but with a shape variability that differs from that of the He~{\sc i} lines. These characteristics suggest that W601 is a PMS He-strong star with He and Si abundance spots on its surface \citep[e.g.,][]{bohlender90}. The observed variations of the He and Si lines on a timescale of one hour (which we interpret as rotational modulation by the abundance spots) are consistent with our predictions of a rotation period shorter than 1.7~d. The line profiles show no systematic, radial velocity variability suggestive of binarity. All the LSD $I$ profiles and Stokes $I$ spectra obtained on the 3 nights show the same small structures in the line profiles. Their origin is unknown, although they are apparently not instrumental.

We plot the LSD Stokes $I$ and $V$ profiles obtained on three different dates in Fig. \ref{fig:w601} (upper frame). We observe some variations in the shape of the $I$ profiles. The red side of the profile has changed between August 2006 and March 2007, while the blue side varied slightly between March 5th and March 8th. The mask used to calculate these profiles contains mainly He~{\sc i} lines. Therefore, the variations in the Stokes $I$ profile reflect primarily those observed in the He~{\sc i} lines of the Stokes $I$ spectrum (Fig. \ref{fig:w601}, lower frame). The polarisation Stokes $V$ profiles, plotted at the top of Fig~ \ref{fig:w601}, show strong and variable Zeeman signatures as broad as the photospheric $I$ profile, and centred on its centroid, which indicates that this signature is created by a magnetic field at the surface of the star. The comparison of the LSD $V$ (upper) to the LSD $N$ (middle) profiles supports that the $V$ signature is real, and not a spurious signature produced by the instrument or data reduction procedure.


The LSD $V$ profiles also vary on a timescale of one hour. Our previous studies of magnetic HAeBe stars have shown that, like the Ap/Bp stars, they host large-scale organised magnetic fields and satisfy the oblique rotator model (Alecian et al. 2008; Folsom etal., in prep.). This model basically consists of a dipole placed inside the star, with a magnetic axis inclined with respect to the rotation axis \citep{landstreet70}. As the star rotates, the configuration of the surface magnetic field seen by an observer changes, and the observed polarisation signal therefore varies. Most Ap/Bp and magnetic HAeBe stars show temporal variations of their polarisation spectra, which are modulated according to the rotation period of the star. By analogy, we propose that the variations of the Stokes $V$ profile of W601 also reflect the rotation of the star. 

Using the Stokes $I$ and $V$ profiles and Eq. (1) of \citet{wade00}, we measured the longitudinal magnetic field $B_{\ell}$ of our data. The values of $B_{\ell}$ are listed in Table 1 and are found to vary between $\sim-1300$ and $\sim1400$~G, indicating that the mean magnetic field modulus is probably larger than 3~kG.

Our data are insufficient to allow us to derive the oblique rotator model parameters of W601. Additional spectropolarimetric observations with dense sampling of the rotation cycle are required to accurately determine the rotation period as well as to model the intensity and the topology of the magnetic field.

%

\section{OI~201 (NGC~2244)}

\begin{figure}[t]
\centering
\includegraphics[width=6.5cm,angle=90]{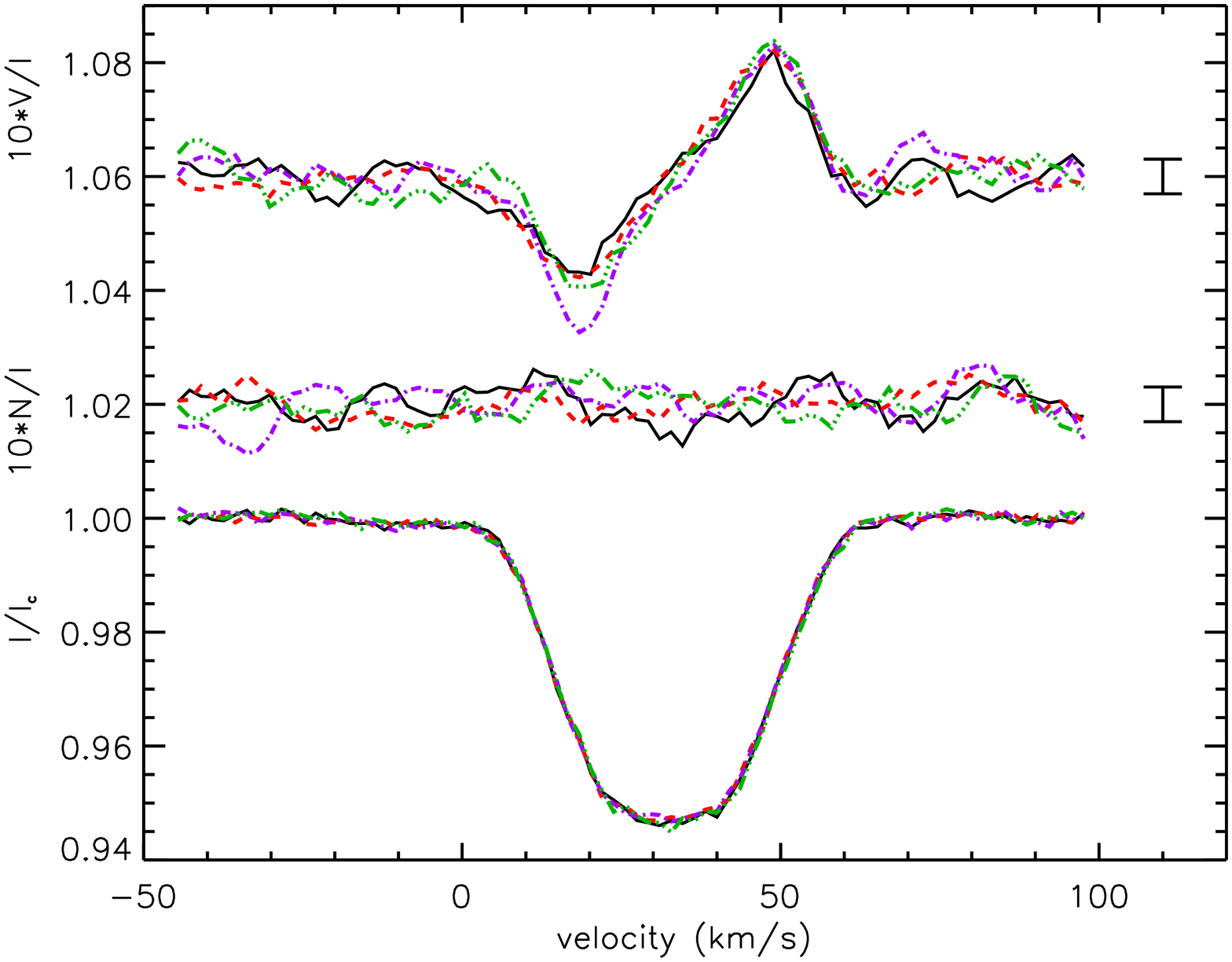}
\hfill
\includegraphics[width=5.6cm,angle=90]{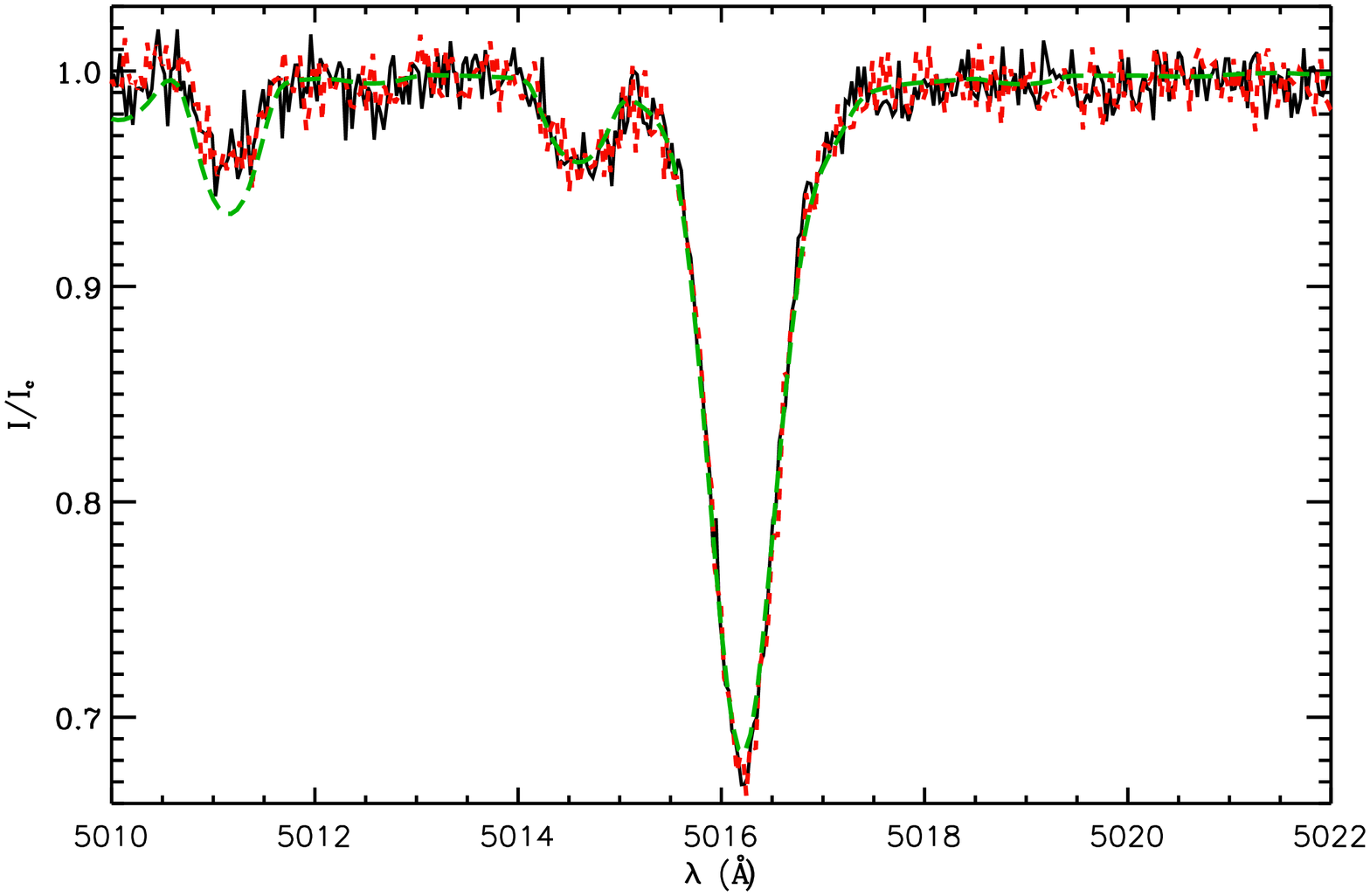}
\caption{{\em Upper frame:}\ LSD $I$ ({\it bottom}), $V$ ({\it upper}) and $N$ ({\it middle}) profiles of OI~201. {\it Full black line}: 4/03, {\it red-dashed line}: 5/03 , {\it blue dot-dashed line}: 6/03, and {\it green dot-dot-dot-dashed line}: 9/03. The mean error bars are plotted next to each profile.  {\em Lower frame:}\ The sharp, nonvariable He~{\sc i}\ $\lambda 5015$ of OI~201, observed on 4/03 ({\it full-black line}) and 9/03 ({\it red-dashed line}). The green long-dashed-line is a TLUSTY/SYNSPEC synthetic profile computed for a solar He abundance.}
\label{fig:oi201}
\end{figure}

The B1 star OI~201 belongs to the cluster NGC 2244, iluminating the Rosette Nebula. The cluster is located at a distance of 1.7~kpc from the sun, with an age of $2.3\pm0.2$~Myr \citep{hensberge00}. 
The proper motion and radial velocity of OI~201 \citep{chen07,perez91}, its reddening law \citep{ducati03}, its position in the HR diagram, and its geographical position in the central blister of the cluster confirm unambiguously that it is a member of NGC 2244. 

OI~201 does not show an IR excess \citep{balog07} and has not been detected as an X-ray source \citep{wang07}. According to \citet{park02}, this star does not have H$\alpha$ in emission, which we confirm with our data. The position of OI~201 in the HR diagram is above, but close to the ZAMS - therefore that star is either an old PMS star or a young MS star. In both cases, this star is very young, but more evolved that W601.

We used the photometric data of \citet{park02} to determine the luminosity of the star: $\log(L/L_{\odot})=4.05\pm0.08$, while its temperature has been determined by \citet{vrancken97}: $T_{\rm eff}=27300\pm1000$~K. We compared the HR diagram position of OI~201 with CESAM PMS evolutionary tracks. We find $M = 12.1^{+0.6}_{-0.5}$~M$_{\odot}$ and $R = 4.7^{+0.6}_{-0.4}$~M$_{\odot}$, leading to $\log g=4.18\pm0.06$~(cgs), consistent with the spectroscopic value of \citet{vrancken97}. Assuming that OI~201 is PMS, its age since the birthline is estimated to be $0.041^{+0.008}_{-0.024}$~Myr (Fig. \ref{fig:hr}), which is inconsistent with the age of the cluster. Therefore this star is more likely a MS star with an age of $3^{+3}_{-3}$~Myr, estimated using CESAM MS evolutionary tracks.

The four spectra, all obtained during a single run of 5 days, show no evidence of emission, either in the Balmer lines or in the He~{\sc i} and O~{\sc i} lines often observed in emission in PMS stars. We observe no variability in any lines and we find that a TLUSTY/SYNSPEC synthetic spectrum with $T_{\rm eff}=25000$~K, $\log g=4.25$~(cgs) and solar abundances, broadened with ${\rm v}\sin i = 23.5\pm0.5$~km.s$^{-1}$, reproduces well the entire spectrum of OI~201 (Fig. \ref{fig:oi201} lower frame). Figure \ref{fig:oi201} (upper frame) shows the LSD profiles of OI~201 observed on 4 different dates. The four Stokes $I$ profiles are perfectly superimposed, illustrating the lack of variability in the spectrum. No structures are visible in the line profiles similar to those detected in W601.


The Stokes $V$ polarisation profile shows a strong Zeeman signature as broad as the $I$ profile, and centred on its centroid, which indicates that this signature is created by a magnetic field with an important poloidal component at the surface of the star. We note that the LSD $N$ profile is totally flat compared to the Stokes $V$ profile, indicating that the Stokes $V$ signature is real.


The Stokes $V$ profiles of the four spectra observed between March 3rd and March 8th have not varied, and yield similar measurements of the longitudinal magnetic field found around -500~G (Table \ref{tab:log}). According to the oblique rotator model described in Sect. 3, the lack of variation in the Stokes $V$ profiles indicates that either the rotation period of the star is very long, the rotation axis of the star is nearly parallel to our line-of-sight, or the magnetic axis is nearly aligned with the rotation axis. With our determinations of $v\sin i$ and $R$, we find that the maximum rotation period of OI~201 is 10.1~d. Therefore a very long period ($\sim$months) is not consistent with the $v\sin i$ and radius. On the other hand, if the inclination of the rotation axis $i$ to the line-of-sight is very small (say lower than 10$^{\circ}$), rotational modulation would be very weak, and perhaps not detectable. Such a pole-on scenario would also imply a rotation period similar to that of W601 (for $i=10\degr$, $v_{\rm eq}=135$~km/s and $P_{\rm rot}=1.8$~d). On the other hand, such a small value of $i$ is statistically very unlikely (less than 1\%). Furthermore OI~201 is the second star of our Herbig survey that shows no variability in the V profile \citep[the first one being HD~190073;][]{catala07}. The probability to observe two pole-on stars in a same sample is small. A final option is that OI~201 has its magnetic axis aligned with the rotation axis, which has been observed in some MS He-strong stars \citep[e.g.,][]{bohlender87}. Additional spectropolarimetric observations are required to distinguish between these different geometries.


%

\section{Discussion}

We have discovered strong magnetic fields at the surfaces of two hot, very young massive stars, both of which are members of young open clusters. Although these stars have similar masses and are both magnetic, they show important physical differences: W601 is a rapid rotator, exhibits strong and variable lines of He and Si, and is situtated in the first half of the PMS phase. OI~201 appears to be a slow rotator, shows normal photospheric chemistry, and may already be on the MS.

The presence of chemical peculiarities in the atmosphere of W601 demonstrates for the first time that such anomalies can develop very early during the PMS evolution of a B star. This prompts the question: why are no similar chemical anomalies observed in the atmosphere of OI~201? This star is only about 20\% hotter and more massive than W601, and it hosts a similar strong magnetic field. Moreover, its slower rotation and apparent lack of a circumstellar accretion disc (as suggested by its lack of emission lines and IR excess) would suggest that its atmosphere is likely more stable than that of W601. We suspect that the differences in the photospheric properties of these stars will yield important new constraints on models of chemical fractionation in the atmospheres and winds of B-type stars.

Unless the rotation axis inclination of OI~201 is smaller than about $10\degr$, it certainly rotates much more slowly than W601 and other B-type stars. Could its advanced age, interpreted within the context of magnetic braking, explain the large difference with respect to W601? 

The answers to these questions must await the acquisition of additional spectropolarimetric data. However, it is clear from the observations presented here that W601 and OI~201 represent two remarkable B-type PMS stars that have the potential to contribute considerably to our understanding of magnetism, rotation, and evolution of intermediate-mass stars.

\begin{acknowledgements}
      EA is supported by the Marie Curie FP6 program. GAW acknowledges support from the Natural Science and Engineering Research Council of Canada (NSERC) and the DND Academic Research Programme (ARP).
\end{acknowledgements}

%

\bibliographystyle{aa}
\bibliography{young_letter}

\end{document}